# Randomization Restrictions: Their Impact on Type I Error When Experimenting with Finite Populations

Jonathan J. Chipman*
Division of Biostatistics, Department of Population Health Sciences,
University of Utah
Cancer Biostatistics, Huntsman Cancer Institute, University of Utah

Oleksandr Sverdlov
Early Development Analytics, Novartis Pharmaceuticals Corporation

Diane Uschner
Product Development Data and Statistical Sciences, F. Hoffmann-La Roche

August 29, 2025

### Abstract

Participants in clinical trials are often viewed as a unique, finite population. Yet, statistical analyses often assume that participants were randomly sampled from a larger population. Under Complete Randomization, Randomization-Based Inference (RBI; a finite population inference) and Analysis of Variance (ANOVA; a random sampling inference) provide asymptotically equivalent difference-in-means tests. However, sequentially-enrolling trials typically employ restricted randomization schemes, such as block or Maximum Tolerable Imbalance (MTI) designs, to reduce the chance of chronological treatment imbalances. The impact of these restrictions on RBI and ANOVA concordance is not well understood. With real-world frames of reference, such as rare and ultra-rare diseases, we review full versus random sampling of finite populations and empirically evaluate finite population Type I error when using ANOVA following randomization restrictions. Randomization restrictions strongly impacted ANOVA Type I error, even for trials with 1,000 participants. Properly adjusting for restrictions corrected Type I error. We corrected for block randomization, yet leave open how to correct for MTI designs. More directly, RBI accounts for randomization restrictions while ensuring correct finite population Type I error. Novel contributions are: 1) deepening the understanding and correction of RBI and ANOVA concordance under block and MTI restrictions and 2) using finite populations to estimate the convergence of Type I error to a nominal rate. We discuss the challenge of specifying an estimand's population and reconciling with sampled trial participants.

---
*Oleksandr Sverdlov and Diane Uschner equally contributed as senior authors. Sverdlov provided key conceptual framing of finite populations and their relevance, particularly in rare and ultra-rare populations. Uschner provided key conceptual insights into ANCOVA adjustments, which informed the methodological corrections and remaining gaps presented in the paper. This work was supported by the support and resources from the Center for High Performance Computing at the University of Utah. AI was used to assist in formatting figures and tables.

# 1 Introduction

Patients who enroll in a randomized clinical trial typically differ from the broader target population of interest. Trial participants include all individuals who meet enrollment criteria, are willing to participate, and are identified from participating sites/catchment areas during the enrollment period (Rosenberger et al. 2019). For this reason, randomized trials are frequently conceptualized and argued to be treated as full samples of finite populations (see Ding et al. (2017) for an extensive list of citations). They carry internal validity but may lack external generalization. This conceptualization is important because carefully articulating the trial population is an essential step in defining the trial estimand. The analytic strategy should then support the trial estimand (International Council for Harmonisation of Technical Requirements for Pharmaceuticals for Human use 2019).

Analysis strategies for finite populations operate under the potential outcomes framework, including Randomization-Based Inference (RBI; see Rosenberger et al. (2019) for a historical and practical overview), tests using finite-population variance estimates for treatment differences (Aronow et al. 2014$a$), and the causal bootstrap (Imbens & Menzel 2021$a$). RBI and the causal bootstrap use the random treatment assignment to account for finite population uncertainty. Whereas, finite population variance estimates place a bound upon the covariance between individual potential outcomes. The estimand is the treatment effect and its uncertainty in the finite population (see for example, Uschner et al. (2024)). As a frame of reference, RBI is an acceptable analytic strategy to the Food and Drug Administration (U.S. Department of Health and Human Services Food and Drug Administration 2023); however, in a recent review, it has not been frequently used as the lead analysis among therapeutics newly approved by the Food and Drug Administration (Carter et al. 2024).

Many trials with continuous outcomes use Analysis of Variance (ANOVA) or Analysis of Covariance (ANCOVA). *For each arm*, ANOVA/ANCOVA assumes patients were randomly sampled from a larger "super-population" of individuals who would be eligible, willing, and available. When there are two arms, the equivalent ANOVA term "two-sample $t$-test" makes this explicit. The super-populations are considered to be hypothetically infinite. This framework contrasts a finite population framework which compares potential outcomes from multiple treatment arms *within a single population*. The ANOVA estimand is the treatment effect and its uncertainty when random sampling from arm-specific super-populations of eligible, willing, and available patients. When ANOVA is used on a random sample from a single population (i.e., an "invoked population model"), the analyst ignores the uncertainty due to randomized treatment assignment and assumes an independent and identically distributed random sample could have been drawn for each arm (Rosenberger et al. 2019).

The inferential consequences of performing RBI when participants are randomly sampled from a super-population are known. The first-order estimate of the sample is unbiased for the broader population (Ding et al. 2017), whereas tests of second-order measures of uncertainty limit conclusions to only trial participants. Hence, conclusions regarding hypothesis testing are limited to only those participants in the study. Given this limitation, RBI maintains correct Type I error and accounts for any design-based restrictions upon randomization. The causal bootstrap accounts for both sampling and treatment assignment uncertainty, thus allowing for broader generalizability while also accounting for any restrictions due to randomization.

There is a close connection between ANOVA and RBI. The two analytic strategies provide asymptotically equivalent tests of the difference of treatment means. Albeit, ANOVA would generalize results to hypothetical super-populations of trial participants, while RBI would limit generalization to trial participants. This gives a degree of reassurance that ANOVA/ANCOVA could be used regardless of whether trial participants are a full, finite population or are randomly sampled from treatment-arm super-populations. Results for asymptotic equivalence typically begin with an assumption of Complete Randomization in which a pre-specified treatment allocation is achieved (e.g., finite population Central Limit Theorems (Li & Ding 2017)).

To clarify terms, the Design of Experiments literature frequently uses *Complete Randomization* to describe a design in which *n* subjects are assigned to treatment groups according to a fixed allocation ratio. However, some randomized clinical trial literature (e.g., Rosenberger & Lachin 2015, Sec. 3.1) call this procedure the *Random Allocation Rule* and use the term *Complete Randomization* to describe a process where each participant is assigned independently, typically using a coin flip—what is more precisely known as *Simple Randomization*. In this paper, we adopt the following terminology: *Simple Randomization* refers to independent, sequential assignment of participants (e.g., by coin flip), and *Complete Randomization* refers to a design where an exact number of participants are assigned to each treatment. *Equal Randomization* is a special case of *Complete Randomization* when treatment allocation is the same for all treatment arms.

Despite the RBI and ANOVA concordance under Complete Randomization, sequentially enrolling trials seldom implement Complete Randomization. Complete Randomization requires exact enrollment of a pre-specified number of participants and may allow for large allocation imbalances throughout the trial. This may be impractical and/or open the trial to chronological bias. Randomly permuted blocks and Maximum Tolerable Imbalance (MTI) randomization schemes are often used to reduce the risk of chance allocation imbalances throughout the study (Berger et al. 2021). Strategies that use an MTI procedure allow for treatment allocation imbalances but within the pre-specified MTI threshold (for example, an imbalance no greater than 4). Soares & Jeff Wu (1983) introduced the Big Stick Design which implements Simple Randomization (i.e., a coin flip) unless forced to assign treatment to maintain the MTI constraint. MTI strategies have appeal for being less predictable than fixed and permuted block randomization, which forces deterministic treatment assignment at regular intervals (Berger et al. 2016).

Although subtle, randomization restrictions alter the RBI randomization distribution, and it is unclear whether and how these common restrictions for chronological imbalance could impact the asymptotic concordance between RBI and ANOVA. In contrast, covariate-adjusted randomization can remove noise from the randomization distribution and more clearly impact RBI and ANOVA concordance (see for example Chipman et al. (2023)).

In this paper, we address the question: What impact do randomization restrictions have on Type I error upon a fully-sampled finite population versus a randomly-sampled yet finite population when performing a finite versus super-population analysis? Empirical simulations account for treatment assignment and random sampling uncertainty in estimating the distribution of ANOVA Type I error across finite populations of the same size and outcome distribution. For simplicity, we focus on outcomes that are drawn from a standard normal distribution and in which there is no treatment effect for any participants (i.e., a "sharp" null hypothesis). However, in Supplemental Materials, we also consider two versions of the weak null for which, on average, there is no treatment effect. We consider Simple, Complete (with equal allocation), Fixed Block, and Big Stick Design Randomization schemes. We focus on comparing Type I error control rather than secondary questions of power. Broadly, our general conclusion is that while Complete and Simple Randomization analyzed with ANOVA maintain strong Type I error control even for small populations, randomization restrictions can strongly impact the Type I error observed in a given trial population. In these simulations, random sampling reduced but did not eliminate the impact. For the sharp null, Type I error control could be corrected by properly adjusting for the randomization restrictions in an ANCOVA model. The following table highlights what is known and what we contribute to understanding the concordance between sampling

and analysis frameworks:

Table 1: Comparison of means using RBI and ANOVA under full versus random samplings of a finite population, and the contributions of this paper.

|  |  | Inference | |
|---|---|---|---|
|  |  | **Randomization-Based Inference (RBI)** | **Analysis of Variance (ANOVA)** |
| **Sampling** | **Full sample** $(n = N)$ | ✓ Consistent with sampling<br>• Estimation: Unbiased<br>• Uncertainty:<br>1. Exactly controlled Type I error<br>2. Generalizes to observed sample | ✗ Not consistent with sampling<br>• Estimation: Unbiased<br>• Uncertainty:<br>1. At least asymptotically correct Type I error*<br>2. Extrapolates to super-population reflective of sample |
|  | **Random sample** $(n < N < \infty)$ | ✗ Not consistent with sampling<br>• Estimation: Unbiased<br>• Uncertainty:<br>1. Exactly controlled Type I error<br>2. Generalizes to observed sample only | ✓ Asymptotically consistent with sampling as $N$ increases<br>• Estimation: Unbiased<br>• Uncertainty:<br>1. At least asymptotically correct Type I error<br>2. Generalizes to super-population reflective of sample |

\* Contribution of this paper: Randomization restrictions slow the rate of Type I error convergence, but adjusting for restrictions (e.g., with ANCOVA) can improve convergence.

For a real-world point of reference on finite populations, Section 2 describes patient horizons and how it can be used to assess trial generalizability in finite populations. We highlight settings with an enumerable patient horizon, including rare and ultra-rare diseases, implementation science trials, and decentralized trials. Section 3 specifies the potential outcomes and sampling framework we're using for hypothesis testing and evaluating Type I error in finite populations. Section 4 provides simulations and empirically-based results for the Type I error across populations of the same size following different randomization schemes. Section 5 provides conclusions and implications. A brief discussion is provided in Section 6 on ways to bridge the gap that can exist when specifying the estimand's target population while acknowledging the trial's sampling frame. The main manuscript compares RBI, ANOVA, and ANCOVA testing, and the supplement further considers testing using finite-population variance estimates.

## 2 Patient horizons

The first introductory paragraph provides an argument that randomized trials should be considered finite populations. Extrapolating beyond the trial participants requires assumptions that either (1) the sample was a random sample of a hypothetical population of similar and independent participants and/or (2) outcomes would be independent and identically distributed to trial participants regardless of where, when, and which (willing and eligible or not) individuals were enrolled. The patient horizon quantifies how many of the patient population could benefit from a treatment (Anscombe 1963) and could be used as an indicator for generalizability assumptions. As the patient horizon-to-trial sample size ratio increases, so too may the required assumptions for real-world generalizability.

In practice, the target patient horizon may not be practically enumerable. Yet, in some instances, the patient horizon may be enumerable and small — or rather, the patient horizon may be small relative to settings with practically non-enumerable populations. The following three real-world settings may draw a full, random, or convenience sample on an enumerable population.

*Rare and ultra-rare diseases*: In the United States, the patient horizon in rare diseases is no more than 200,000 individuals (U.S. Food and Drug Administration 2024). Applying the England and Scotland criteria for ultra-rare diseases, the United States patient horizon is approximately 6,600 individuals (Michaeli et al. 2023). Many trials of rare and ultra-rare diseases enroll patients who belong to a registry of known diagnoses.

*Implementation science*: Implementation science trials often test for the effectiveness of an intervention in small, finite populations (such as within 1-3 health systems). Later stage implementation trials (e.g., Hybrid Type 3 effectiveness trials) may study strategies to upscale the intervention in a broader, possibly non-enumerable population (Bauer et al.

2015).

In an analogous fashion, most Phase I-III trials may be viewed as establishing efficacy (not effectiveness) in small, finite populations. Phase IV trials then look to assess effectiveness in broader populations.

*Decentralized trials*: Decentralized trials are characterized by having "trial-related activities occur at locations other than traditional clinical trial sites" (U.S. Food and Drug Administration 2023) including how the trial is conducted and data are managed (Chen et al. 2025). Many decentralized trials enroll their patients from an enumerable registry of patients, for example, the Scale Up II trial, which tested multiple interventions to increase testing for COVID-19 (Del Fiol et al. 2024). While not all decentralized trials may be large-scale, pragmatic trials, many have potential to reach a broader patient population.

Setting aside eligibility criteria, willingness and availability are often hidden inclusion/exclusion criteria. The impact of willingness and availability could be context-dependent. For example, individuals with a rare or ultra-rare disease could be more willing to participate in a randomized trial. Willingness can be partially assessed by the number of patients who were screened, met the inclusion and exclusion criteria, yet declined participation. The CONSORT 2025 flow diagram requires this information (Hopewell et al. 2025). Additionally, though geography may not alter the untreated disease course, varying clinical practices across regions and countries may influence the outcome.

In the forthcoming simulations, we will assess scenarios where the sample is the full population and where it is a random sample from a broader, finite patient horizon. Although random sampling in the latter scenario may be questionable, rather than convenience sampling, we will consider this an ideal scenario.

# 3 Hypothesis testing in finite populations

Suppose a finite population exists with $i = 1, \ldots, N$ participants and that a two-armed randomized trial is implemented having $n \leq N$ participants with $n_0 = n_1 = n/2$ participants on each arm. For each individual, $R_i \in \{0 = no, 1 = yes\}$ indicates trial participation and $\mathbf{R} := (R_1, \ldots, R_N)^\top$ is a vector of which individuals participated in the trial. Treatment assignment is denoted as $Z_i \in \{\emptyset, 0, 1\}$ which takes on the empty set when $R_i = 0$ and 0 or 1 otherwise. Define $\mathbf{Z} := (Z_1, \ldots, Z_N)^\top$ as the vector of participating individuals' treatment assignments. When $n < N$, uncertainty can be attributed to the stochastic nature of sampling participants ($\mathbf{R}$) and the stochastic nature of treatment assignment ($\mathbf{Z}$). To put this in context, ANOVA and ANCOVA account for $\mathbf{R}$ but not $\mathbf{Z}$ in estimating uncertainty, and RBI estimates uncertainty accounting for $\mathbf{Z}$ but not $\mathbf{R}$. See Imbens & Menzel (2021*a*) for more context accounting for $\mathbf{R}$ and $\mathbf{Z}$ uncertainty.

Individuals have potential outcomes $Y_i(0)$ and $Y_i(1)$, respectively when $z_i = 0$ and $z_i = 1$. Under the sharp null hypothesis $Y_i(0) = Y_i(1)$. However, the null hypothesis could be relaxed so long as it satisfies $\bar{Y}(0) = \bar{Y}(1)$. Let $\mathbf{Y(0)} := (Y_1(0), \ldots, Y_N(0))$ and $\mathbf{Y(1)} := (Y_1(1), \ldots, Y_N(1))$. Define $\mathbf{Z}_{obs} \in \mathbf{Z}$ as the observed randomization sequence. The quantity of interest is the finite population average treatment effect

$$\tau_{ATE} = \frac{1}{N} \sum_{i=1}^{N} (Y_i(1) - Y_i(0))$$

which can be estimated as

$$\hat{\tau}_{ATE} = \frac{1}{n_1} \sum_{i: R_i = 1} (Z_i Y_i(1)) - \frac{1}{n_0} \sum_{i: R_i = 1} ((1 - Z_i) Y_i(0)).$$

For a given randomization scheme, $RS$, and trial sample size, $n$, $\mathbf{Z}$ has a probability distribution on the set $\Omega_{RS} := \{0, 1\}^n$. Under the null hypothesis, an exact test can be implemented on the observed summary statistic, $S(\mathbf{Z}_{obs}, \mathbf{Y}(1), \mathbf{Y}(0) \mid \mathbf{R}) = \hat{\tau}_{ATE}$, as follows, with $1(\cdot)$ being the indicator function:

$$p_{RBI} = \sum_{Z_j \in \Omega_{RS}} 1\left[|S(Z_j, Y(1), Y(0) \mid R)| \geq |S(Z_{obs}, Y(1), Y(0) \mid R)|\right] \cdot P(Z_j \mid \Omega_{RS}).$$

In practice, it is often impractical to fully enumerate $\Omega_{RS}$. Yet, the randomization test can be implemented with sufficient precision by generating $L$ randomization sequences from $\Omega_{RS}$ and estimating

$$\hat{p}_{RBI} = \frac{1}{L} \sum_{l=1}^{L} 1\left[|S(Z_l, Y(1), Y(0) \mid R)| \geq |S(Z_{obs}, Y(1), Y(0) \mid R)|\right].$$

$L$ can be chosen based on the precision to estimate the p-value. For example, if p $= 0.05$, an $L$ of 10,000 estimates $\hat{p}_{RBI}$ within $+/-$ 0.004 using a 95% binomial confidence interval. A Type I error occurs when $p_{RBI} \leq \alpha$ (or $\hat{p}_{RBI} \leq \alpha$) under the null hypothesis. Since $\mathbf{Z}$ is fully enumerated and forms the distribution of the test statistic, exactly $(\alpha \cdot 100)\%$ of $\mathbf{Z}$ will result in a Type I error under the null hypothesis. It is the proportion of randomization sequences for which significance would be declared.

Although ANOVA and ANCOVA would be discordant with the sampling frame when $n = N$, that does not preclude the analyst from using ANOVA or ANCOVA. In the context of a linear model $Y_i = \tau_{ATE} + \boldsymbol{\beta}^\top \mathbf{x}_i + e_i$, where $\boldsymbol{\beta} \in \Re^{p+1}$ is a vector of coefficients corresponding with $\mathbf{x}_i = (1, x_{i1}, \ldots, x_{ip})^\top$ representing an intercept and adjusting covariates and $e_i$ is an independent residual of the randomly-sampled individual's outcome from the population predicted outcome. In ANOVA, $p = 0$ so that the model only includes the intercept $\beta_0$. Define $\hat{\tau}_{ANOVA}$ and $\hat{\tau}_{ANCOVA}$ as the estimate of $\tau_{ATE}$ after respectivelly fitting an ANOVA or ANCOVA model. The ANOVA linear model estimate, $\hat{\tau}_{ANOVA}$ is equivalent to the unadjusted difference in means $\hat{\tau}_{ATE}$. In our focus here, ANCOVA adjusts for $p$ covariates that correspond with randomization restrictions from Complete Randomization. For example, it could denote membership of a given block when implementing fixed or permuted block randomization or center if randomization occurred within each center of a multi-center trial. From their respective models, the sampling uncertainty in estimating $\hat{\tau}_{ANOVA}$ and $\hat{\tau}_{ANCOVA}$ can be estimated by their standard error – $\hat{SE}_{ANOVA}$ and $\hat{SE}_{ANCOVA}$. Asymptotically, the t-test for the coefficient $\hat{\tau}_{ANOVA}$ and $\hat{\tau}_{ANCOVA}$ follows a $t$-distribution with $n-2$ and $n-p-2$ degrees of freedom. Statistical significance under ANOVA and ANCOVA is:

$$p_{ANOVA} = t^{-1}\left(\frac{\hat{\tau}_{ANOVA}}{\hat{SE}_{ANOVA}}, n-2 \mid Z\right) \text{ and}$$

$$p_{ANCOVA} = t^{-1}\left(\frac{\hat{\tau}_{ANCOVA}}{\hat{SE}_{ANCOVA}}, n-p-2 \mid Z\right)$$

where $t^{-1}$ is the cumulative distribution function of the $t$-distribution.

In a finite population, where $n = N$, ANOVA and ANCOVA Type I error is estimated similarly as with RBI – as the proportion of times significance would have been declared across all possible randomization sequences. This is because potential outcomes remain fixed and uncertainty is attributed to the possible randomization sequences. In contrast to RBI, there is no guarantee of an exact Type I error because ANOVA and ANCOVA accounts for **R** but not **Z** uncertainty.

Sampling $n < N$, introduces sampling uncertainty. Type I error for the population is the proportion of times a test performed on a random sample *and* across the possible treatment assignments within the random sample, would reject the null hypothesis.

We define Type I error convergence as occurring when all populations of a given size, $N$, have the same Type I error within some error. As an empirical measure of convergence, we compared the difference between the 2.5 and 97.5 percentiles of Type I error estimates from populations of size $N$.

# 4  Empirically evaluating Type I error

Through simulations, we now consider the impact of randomization restrictions upon Type I error across finite populations of the same size and outcome when accounting for treatment assignment and sampling uncertainty. The anticipation is that the Type I error of using ANOVA will asymptotically converge in probability to 0.05 but that convergence rates may depend upon randomization restrictions. An overview of the simulations is provided in Table 2.

---

**Step 1: Generate a population of size $N$ with independent draws from $\mathcal{N}(0,1)$**

A) If $n < N$, sample $n$ without replacement. Skip step if $n = N$.

B) Generate a randomization sequence from the randomization scheme.

C) Perform tests and record whether the test would reject ($p_{test} < \alpha = 0.05$):

– ANOVA and ANCOVA adjusting for randomization restrictions and

– Difference in means test using finite population variance estimates.

D) Repeat B) and C) `nrands` times to estimate the within-sample rejection rate across randomization sequences.

E) Repeat A) through D) `nsamps` times (if $n < N$) to estimate the rejection rate across samples.

F) Estimate Type I error, per test, for the population as the proportion of total rejections.

**Step 2: Repeat Step 1 npops times**

---

Table 2: The algorithm used to estimate Type I error within a finite population and across finite populations of the same size and outcome distribution. When $n = N$, we used `nrands` = 10,000 and `npops` = 30,000. For computational feasibility, we used `nrands` = 1,000, `nsamps` = 100, and `npops` = 10,000 when $n < N$.

Full sampling of a finite population ($n = N$): Data for `npops` = 30,000 finite populations were generated for sample sizes of $2^{4:10}$ and midpoints between $2^{4:7}$, which based upon simulations was sufficient to remove monte-carlo noise in estimating Type I error. Observed outcomes were drawn from a standard normal distribution, and the sharp null was assumed. The Supplemental Material expands results for two non-sharp null hypotheses. Randomization schemes included Simple Randomization, Complete Randomization (with $n_0 = n_1$), Fixed Block Design Randomization with blocks of size 2, 4, 6, and 8, and Big Stick Design Randomization with an MTI of 2, 3, and 4. $L=$`nrands`$=10,000$ randomization sequences were generated for each randomization scheme to estimate finite population Type I error within +/- 0.004. For each randomization sequence, we recorded statistical significance following ANOVA and ANCOVA adjusted for randomization restrictions with $\alpha = 0.05$. We estimated the Type I error for each of the 30,000 finite populations and obtained an empirical distribution of Type I error for a population of size $N$. Finite population tests for the difference in means, using "Neyman" and "Aronow" standard errors, are provided in the Supplemental Material.

For ANCOVA design adjustments, we adjusted Fixed Block Designs for the block to which each participant belongs. For the Big Stick Design we considered four adjustments: an indicator for being at an MTI threshold (-1 for -MTI, 1 for MTI, and 0 otherwise), an indicator of the current imbalance level (ranging from -MTI to MTI by one), indicators consistent with using a fixed block design with blocks size half the size of the MTI threshold, and indicators consistent with using a fixed block design with blocks of size two. Altogether, the adjustments for Fixed Block and Big Stick Design Randomization aimed to isolate observations within which Simple or Complete Randomization occurred.

Sampling from a finite population ($n < N < \infty$): For sample sizes of 20, 50, and 200, data for `npops`$=30,000$ finite populations were generated where the patient horizon was a multiplicative factor of the sample size. We used $N/n$ of 1.0, 1.1, 1.2, 1.5, 2, 4, 10, 50, and 100. These settings aimed to create plausible scenarios of performing a randomized trial on rare or ultra-rare populations (for example a trial with $n = 200$ and $N = 200,000$). Subsequently informed by these simulations, data for 30,000 finite populations were generated for populations with finite patient horizons of $N = 1.1 * n$ where $n = 2^{4:10}$ the midpoints between $2^{4:7}$. Sharp null outcomes, randomization schemes, and statistical tests were generated/performed as described in simulations when $n = N$. For computational feasibility and to capture both sampling and randomization uncertainty, we set $L =$ `nrands` $= 1,000$ for each of `nsamps` $= 100$ samples from a given population.

When sampling $n = N$, we used $L$ and a 95% binomial confidence interval to inform the expected monte-carlo error in estimating Type I error. However, for sampling $n < N$, we use Simple and Complete Randomization as a convergence frame of reference.

For both $n = N$ and $n < N$, we provide ANOVA and ANCOVA comparisons to the theoretical error rate of 5% for RBI using a significance level of alpha $= 0.05$.

## 4.1  Randomizing the full population ($n = N$)

In these simulations, ANOVA Type I error converged to 0.05 within +/- 0.004 most quickly under Simple Randomization and Complete Randomization (Figure 1). On average, the Type I error rate across finite populations was 0.05 when using ANOVA or ANCOVA, regardless of the randomization scheme and sample size. However, unlike RBI, which ensures exact Type I error for all finite populations, the ANOVA Type I error rate was not guaranteed to be 0.05 for any single finite population.

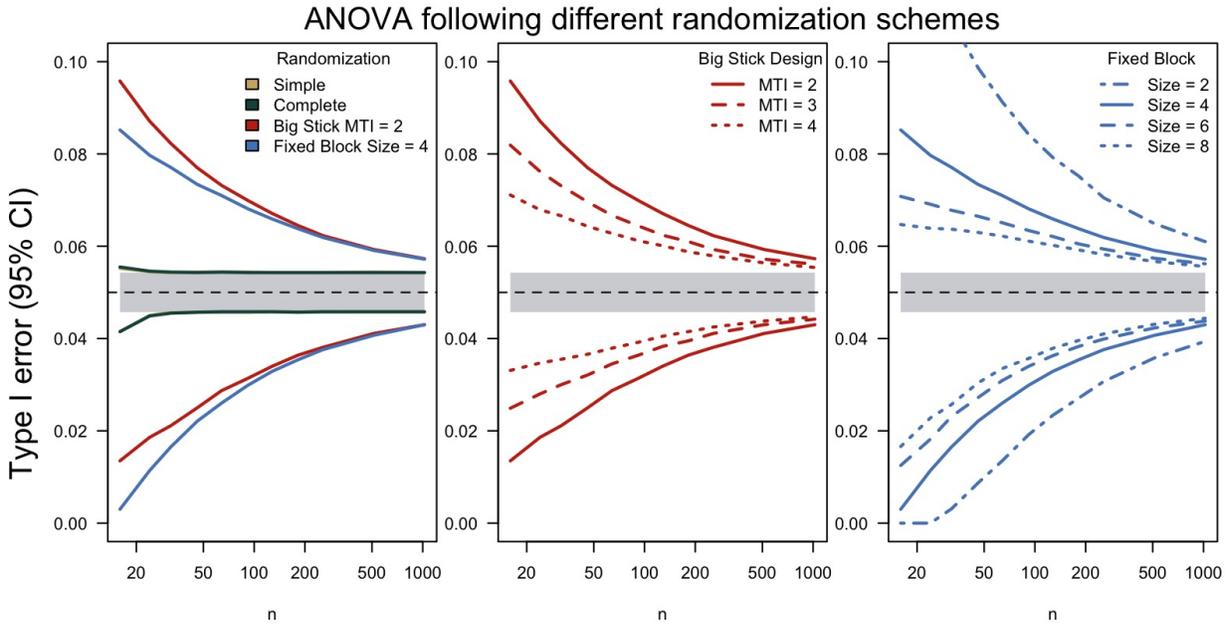

Figure 1: Type I error across populations when $n = N$ and following different randomization schemes. The $2.5^{th}$ and $97.5^{th}$ percentiles are provided as 95% empirical confidence intervals. Type I error convergence worsened with increased restrictions for Big Stick Design (middle figure) and Fixed Block (right figure) randomization schemes. The grey band reflects monte-carlo error when estimating statistical significance with $L$=10,000. The dashed line at 0.05 is the theoretical Type I error under RBI which, by design, is controlled exactly for all populations and randomization schemes. (Simple Randomization appears hidden behind Complete Randomization.)

By $n = 32$, 95% of finite populations under Simple and Complete Randomization had a Type I error rate within the monte-carlo bounds (see expanded results in the Supplemental Material). This is the expected amount to be within bounds that were based upon 95% binomial confidence intervals for a 0.05 Type I error with $L = 10,000$. However, increasing randomization restrictions decreased the rate of Type I error convergence. By 1,024 observations only 56% of finite populations under Fixed Blocks of size 2 (the most restrictive of the schemes studied) had a Type I error within the monte-carlo bounds. Convergence rates were roughly similar for the Big Stick Design and Fixed Block Designs when using a common MTI. By 1,024 observations 76% of finite populations had a Type I error within the monte carlo bounds when using the Big Stick Design with an MTI of 2 and Fixed Block Design with blocks of size 4.

Using ANCOVA to adjust for blocks quickly improved the Type I error convergence rate for Fixed Block Designs (Figure 2). For blocks of size 4, the Type I error convergence rate was roughly similar to ANOVA under Simple or Complete allocation in trials by about $n = 64$. However, none of the ANCOVA adjustments for the Big Stick Design improved the Type I error convergence as quickly as block adjustment did for Fixed Block Design. There were still moderate improvements, yet for a Big Stick Design with an MTI of 2, the percent of populations within monte-carlo bounds ranged from 60% to 88% for sample sizes between 128 and 1,024.

In the Supplemental Material, we explored two non-sharp null scenarios – one with a potential outcome set to zero for all participants and another with potential outcomes being a separate, independent draw from the standard normal distribution. In both cases, the average Type I error across populations remained at 0.05; however, the ANOVA and ANCOVA Type I error across simulated finite populations was highly skewed (unlike under the sharp null). Under the sharp null, we also considered using Neyman and Aronow finite population variances for the difference in means. Type I error was increased under the sharp null hypothesis but converged to expected error rates as sample size increased.

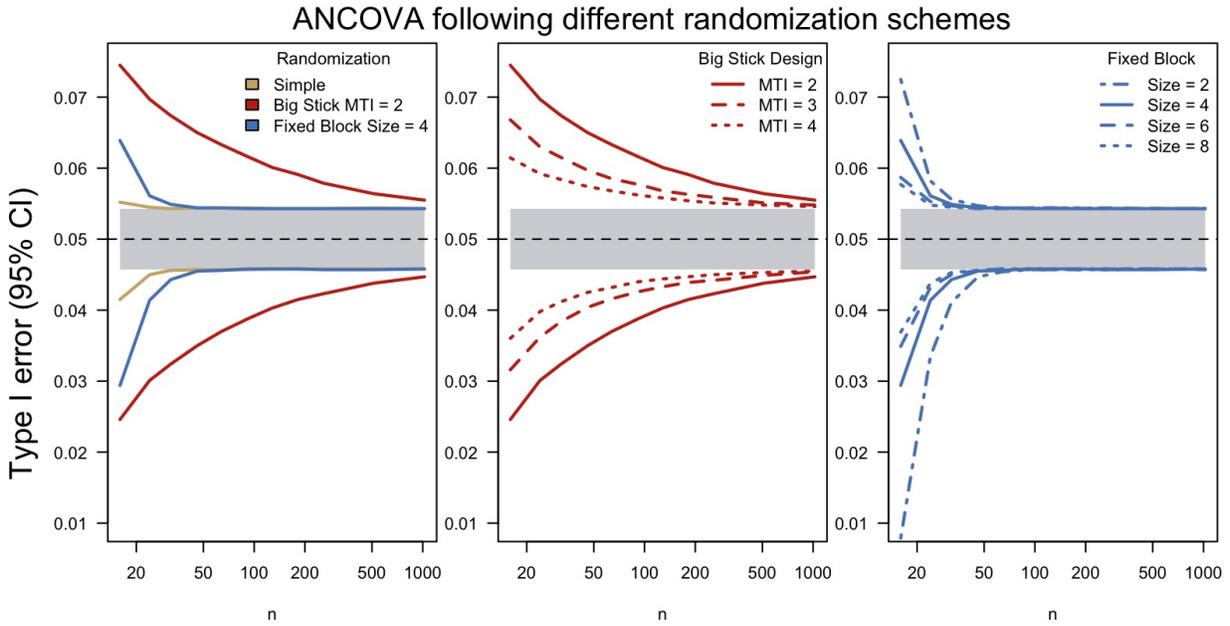

Figure 2: Type I error convergence rates when using ANCOVA to adjust for randomization restrictions following different randomization schemes (dashed lines). ANOVA under each restricted randomization scheme and Simple Randomization are provided as a frame of reference (solid lines). ANCOVA quickly improved Fixed Block Type I error convergence to be similar to ANOVA under Simple Randomization.

## 4.2 Randomizing a sample of the population (n < N)

The smallest ratio of $N$ to $n$ we considered was 110%. Yet, even this relatively small increase markedly improved the ANOVA Type I error convergence following all randomization schemes considered (Figure 3). For all $N/n$ that we considered, the Type I error convergence seemed to plateau for $n = 50$ and $n = 2,000$ and for each randomization scheme.

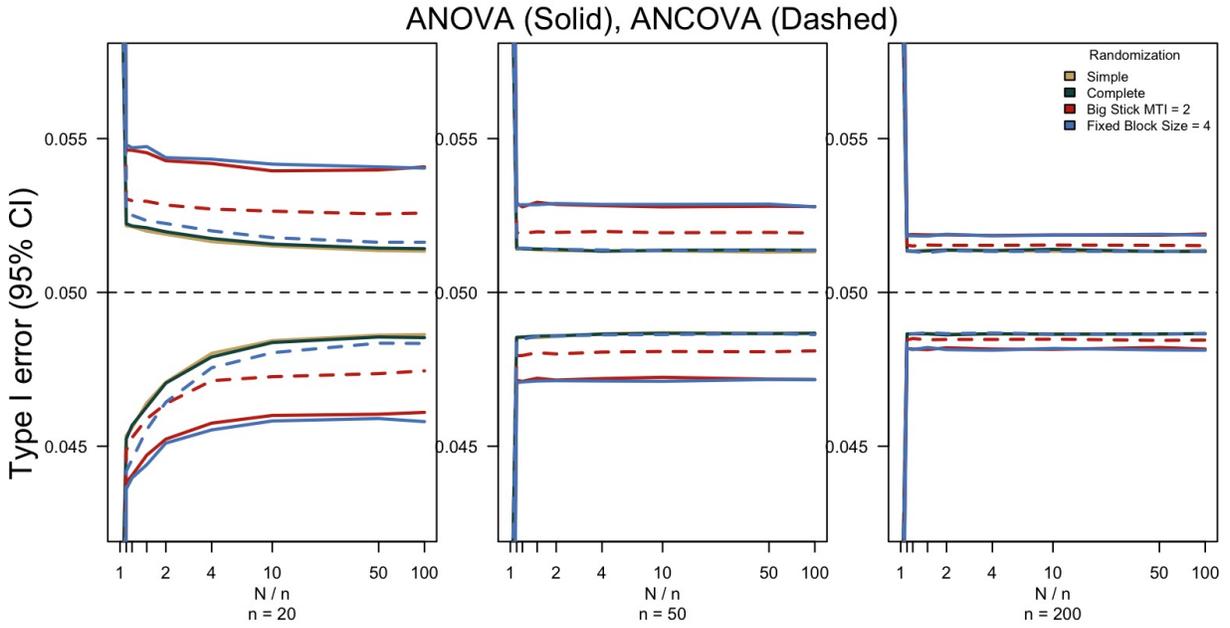

Figure 3: Type I error convergence rates when $n < N$ for $n = 20$, 50, and 200. Ticks denote simulation settings for the population size $N$ relative to $n$. ANCOVA corrects Type I error following Fixed Block Design more so than following Big Stick Design. While this figure captures Type I error, it does not convey that RBI (dashed line at the theoretical Type I error of 0.05) limits conclusions to a finite sample and ANOVA/ANCOVA extrapolates conclusions to a broader super-population.

The more notable benefit to Type I error convergence, relative to Simple and Complete Randomization, was increasing $n$. Still, in these simulations, the ANOVA Type I error following Fixed Block Design (block size 4) and Big Stick Design (MTI = 2) had nearly, but not fully, converged to be equivalent to Simple and Complete Randomization by $n = 1,024$ (Figure 4). By roughly $n = 40$ and beyond, ANCOVA improved Type I error convergence for Fixed Block Design to be similar to Simple and Complete Randomization. As with sampling $n = N$, ANCOVA moderately but not fully improved Type I error convergence following the Big Stick Design.

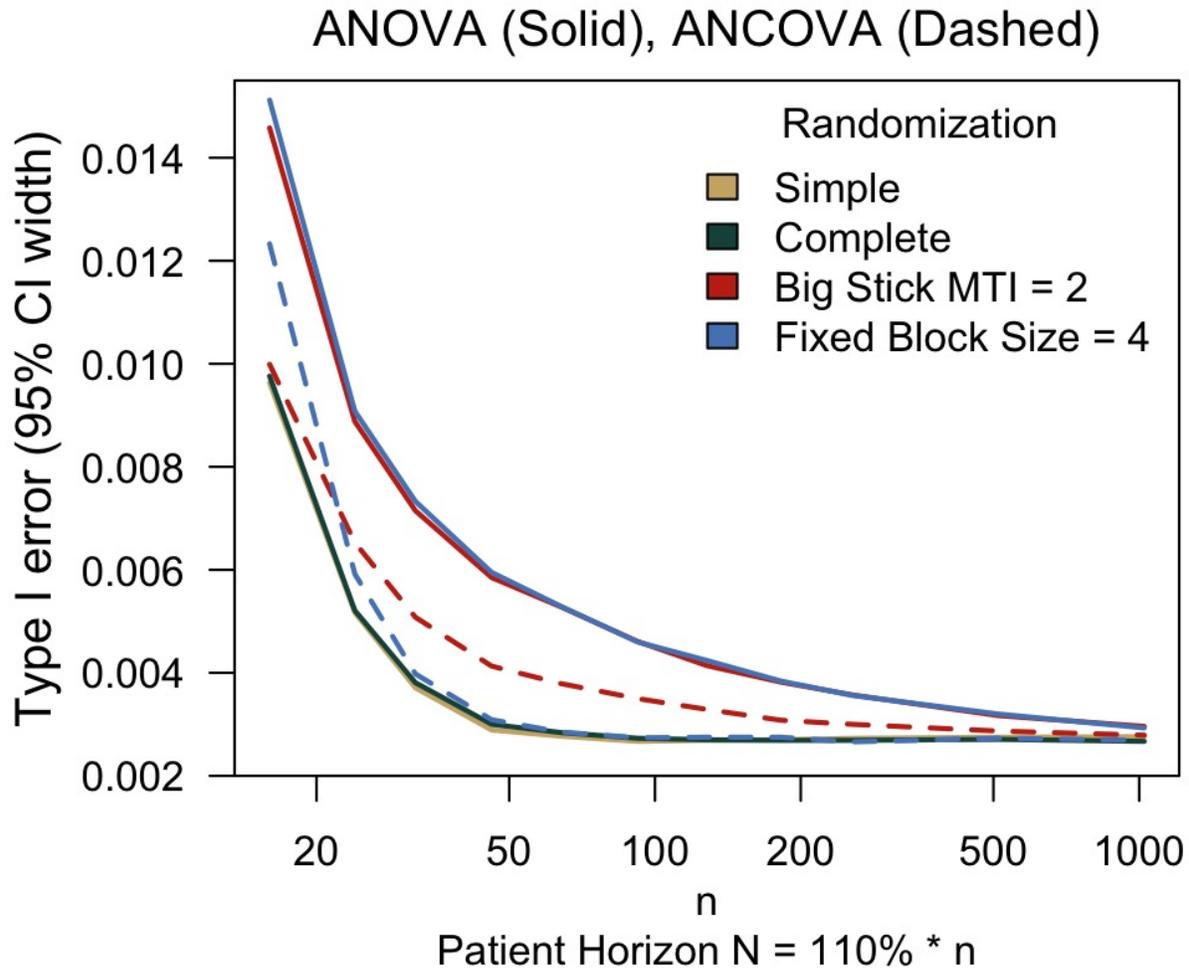

Figure 4: Type I error convergence when sampling from a finite population patient horizon 110% larger than the sample. We focused on $N/n = 110\%$ because the impact of increasing $N/n$ seemed had plateaued for $n = 50$ and $n = 200$ by $N/n = 100\%$ (see Figure 3).

## 5 Conclusion and implications

These empirical results build upon our understanding of the concordance between RBI and ANOVA when sequentially sampling from finite populations and accounting for treatment assignment and sampling uncertainty. Our primary contribution is that the ANOVA Type I error convergence rate in finite populations is impacted by randomization restrictions. More restrictions worsened the Type I error convergence rate. However, this could be corrected by adjusting for randomization restrictions via ANCOVA. We were able to make the correction for Fixed Block designs but leave as open how to do so for the Big Stick Design. Because Complete Randomization occurs within each fixed block, it is not too surprising that Fixed Block ANCOVA quickly corrected for Type I error. This may be analogous to Li & Ding (2017) extending the finite population central limit theorems to cluster-randomized trials with clusters being completely randomized.

These practical implications are made with the view that participants in a randomized trial are drawn from a single population and are generally not similar to a broader population of participants. They are eligible participants who are geographically available at the time of the trial and who are willing to participate. However, when we suppose that trial participants could be a random sample of a broader population, our simulations suggest that adjusting for randomization restrictions remains relevant in small- to moderate- trials.

Though we aimed to account for all aspects of uncertainty, results for $n < N$ were computationally limited in accounting for both treatment assignment and sampling uncertainty (i.e., settings for `nrands` and `nsamps`). For both $n = N$ and $n < N$ conclusions, we aimed to use an idealistic scenario in which outcomes are drawn from a standard normal distribution and no treatment effect exists for any participants. However, we further considered two non-sharp null hypotheses in the Supplemental Material. Many other scenarios of outcome distributions, target treatment allocation, and randomization schemes exist that could also impact the results of the simulations.

## 6 Discussion

An aspect of this work may seem philosophical: Should one consider trial participants to be a full, finite population or a random sample of hypothetical patients similar to them?

We are inclined to view trial participants as coming from a single population and as being a full, finite population. While this leads us to favor RBI, we are aware that RBI is not as well-developed as many model-based strategies. For example, there is some but relatively less work in how to handle missing outcomes with RBI (Ivanova et al. 2022, Heussen et al. 2023). It may be reasonable to make sampling assumptions to address complexities that RBI may not yet be equipped to address. However, all else being equal, we would defer to RBI for its ability to exactly control Type I error and for not relying upon model assumptions.

Regardless of the viewpoint taken, there is often a generalizability gap between the "target population" specified in a clinical trial estimand and the sample of trial participants. To reconcile this gap, we recommend that the estimand distinguish between the target population and the sampled population. The ICH E9 addendum emphasizes the importance of sensitivity analyses to assess deviations from the target estimand; however, we have more often seen this with regards to estimation and quantification of uncertainty. We would like to see additional emphasis given to the sensitivity of "target population" to being reflected by the trial sample. This could be by discussing how susceptible outcomes may be impacted across clinics, geography, and time. And, it could include a reflection of how many potential trial participants were eligible yet declined participation.

# 7 Disclosure statement

The authors have no conflicts of interest.

# 8 Data Availability Statement

The code used for simulations are provided in the supplement.

# 10 Supplemental Material

## S1 Big Stick ANCOVA adjustments

We considered the following design adjustments for ANCOVA:

1. Create an indicator -1 when at -MTI, 1 when at MTI, and 0 otherwise
2. Create an indicator of current imbalance (an integer ranging from -MTI to MTI)
3. Fixed block of size 2 * MTI
4. Fixed block of size 2

Of the design adjustments considered, none quickly improved Type I error convergence to Simple or Complete Randomization as when adjusting for Block Randomization. Making an indicator adjustment of -1 when at -MTI, 1 when at MTI, and 0 otherwise appeared to most quickly improve Type I error across sample sizes explored (Figure S1).

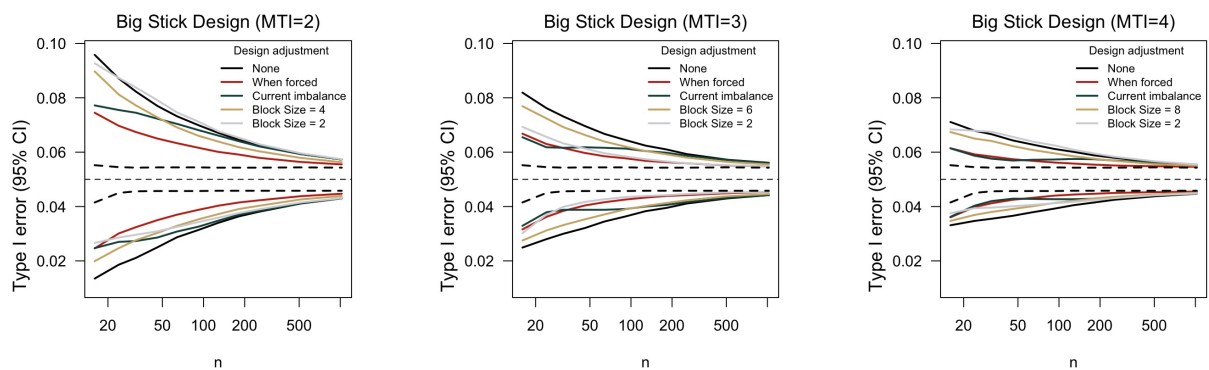

Figure S1: ANCOVA Type I error when performing different design adjustments. ANOVA Type I error under each Big Stick Design (solid black) and under Simple Randomization (dashed black) are provided as a frame of reference.

## S2 Finite population standard errors

In a multi-arm setting, the variance of the finite population difference in means cannot be directly estimated. The problem requires an estimate of the variability under each treatment arm and of the covariance between the individual potential outcomes. However, the covariance cannot be estimated because only one potential outcome is observed.

For the two-arm setting, Neyman provided a conservative estimate of the variance by dropping the non-negative covariance (see (Imbens & Menzel 2021b)). (Note, the estimator is not a pooled estimate of variability; whereas, the results in the main paper use a pooled estimator). Aronow et al. (2014b) uses the marginal distributions of observed outcomes to derive a "sharp" bound on the estimated variance, which yields more precise estimates of the treatment effect (and more narrow confidence intervals).

For the settings in the main paper when $n = N$, we evaluated Wald-Tests using Neyman and Aaronow finite population standard errors. Under the sharp null, Aronow standard errors underestimated the variability and thus provided anti-conservative tests. This was most noticeable under Complete Randomization. However, underestimation decreased with increased sample size. This result for the sharp null was also observed in simulations by Aronow et al. (2014b).

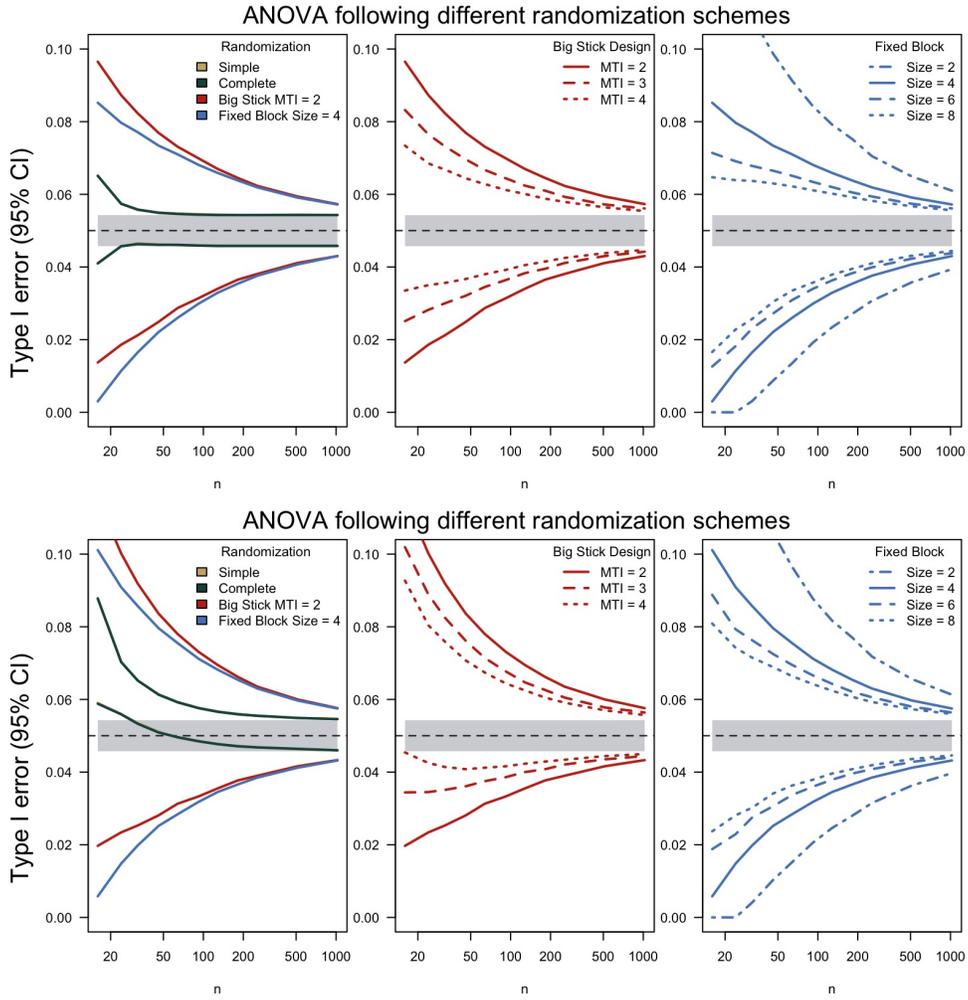

Figure S2: Top: Wald-Test Type I error using Neyman estimates of uncertainty; bottom: using Aronow estimates of uncertainty. The sharp null was used in these simulation settings.

## S3 Sharp and non-sharp nulls

In addition to the sharp null, we considered two non-sharp nulls where:

1. Normal, 0 Null: $Y_i(0) \sim N(0,1)$ and $Y_i(1) = 0$; and

2. Normal, Normal Null: $Y_i(0)$ and $Y_i(1)$ are independent draws from $N(0,1)$.

We then replicated the simulations as described in the manuscript for settings when $n = N$.

For the two non-sharp nulls, the distribution of Type I error was heavily skewed across populations of the same and distribution (Figure S3). The average Type I error across finite populations remained at 0.05 under the Normal, Normal Null (Table S5). However, the Normal, 0 Null had an inflated Type I error for all randomization schemes (including Simple Randomization) until slightly above 100 observations (Table S3).

The skewness remained for increasing sample sizes. For the sharp null, and under Simple Randomization, 95% of finite populations had a Type I error within rounding error (0.05 +/- 1.96 * sqrt(0.05 * 0.95 / 10,000); Table S2). This is the expected target. However, by 100 observations and thereafter, under five percent of finite populations have a Type I error within the expected rounding error (Tables S4 and S6).

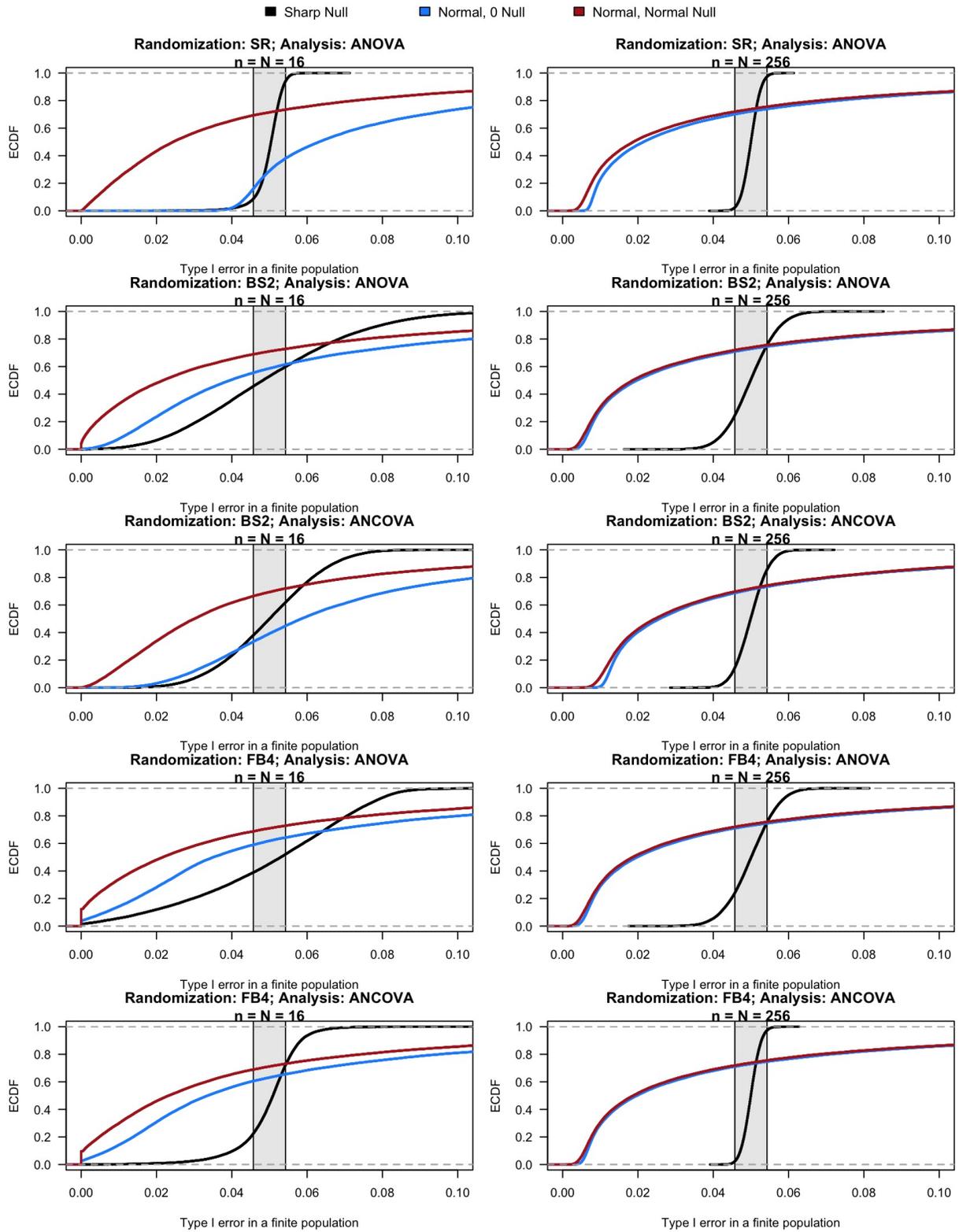

Figure S3: The Empirical Cumulative Distribution Functions for sharp and non-sharp null specifications. The grey bound reflects the rounding error estimating Type I error under 10,000 randomization sequences.

## S3.1 Sharp Null Type I error: Average and convergence ($n = N$)

|           | 16   | 24   | 32   | 46   | 64   | 92   | 128  | 182  | 256  | 512  | 1024 |
|-----------|------|------|------|------|------|------|------|------|------|------|------|
| SR        | 0.05 | 0.05 | 0.05 | 0.05 | 0.05 | 0.05 | 0.05 | 0.05 | 0.05 | 0.05 | 0.05 |
| CR        | 0.05 | 0.05 | 0.05 | 0.05 | 0.05 | 0.05 | 0.05 | 0.05 | 0.05 | 0.05 | 0.05 |
| BS2       | 0.05 | 0.05 | 0.05 | 0.05 | 0.05 | 0.05 | 0.05 | 0.05 | 0.05 | 0.05 | 0.05 |
| BS3       | 0.05 | 0.05 | 0.05 | 0.05 | 0.05 | 0.05 | 0.05 | 0.05 | 0.05 | 0.05 | 0.05 |
| BS4       | 0.05 | 0.05 | 0.05 | 0.05 | 0.05 | 0.05 | 0.05 | 0.05 | 0.05 | 0.05 | 0.05 |
| BS2 adj 1 | 0.05 | 0.05 | 0.05 | 0.05 | 0.05 | 0.05 | 0.05 | 0.05 | 0.05 | 0.05 | 0.05 |
| BS3 adj 1 | 0.05 | 0.05 | 0.05 | 0.05 | 0.05 | 0.05 | 0.05 | 0.05 | 0.05 | 0.05 | 0.05 |
| BS4 adj 1 | 0.05 | 0.05 | 0.05 | 0.05 | 0.05 | 0.05 | 0.05 | 0.05 | 0.05 | 0.05 | 0.05 |
| BS2 adj 2 | 0.05 | 0.05 | 0.05 | 0.05 | 0.05 | 0.05 | 0.05 | 0.05 | 0.05 | 0.05 | 0.05 |
| BS3 adj 2 | 0.05 | 0.05 | 0.05 | 0.05 | 0.05 | 0.05 | 0.05 | 0.05 | 0.05 | 0.05 | 0.05 |
| BS4 adj 2 | 0.05 | 0.05 | 0.05 | 0.05 | 0.05 | 0.05 | 0.05 | 0.05 | 0.05 | 0.05 | 0.05 |
| BS2 adj 3 | 0.05 | 0.05 | 0.05 | 0.05 | 0.05 | 0.05 | 0.05 | 0.05 | 0.05 | 0.05 | 0.05 |
| BS3 adj 3 | 0.05 | 0.05 | 0.05 | 0.05 | 0.05 | 0.05 | 0.05 | 0.05 | 0.05 | 0.05 | 0.05 |
| BS4 adj 3 | 0.05 | 0.05 | 0.05 | 0.05 | 0.05 | 0.05 | 0.05 | 0.05 | 0.05 | 0.05 | 0.05 |
| BS2 adj 4 | 0.05 | 0.05 | 0.05 | 0.05 | 0.05 | 0.05 | 0.05 | 0.05 | 0.05 | 0.05 | 0.05 |
| BS3 adj 4 | 0.05 | 0.05 | 0.05 | 0.05 | 0.05 | 0.05 | 0.05 | 0.05 | 0.05 | 0.05 | 0.05 |
| BS4 adj 4 | 0.05 | 0.05 | 0.05 | 0.05 | 0.05 | 0.05 | 0.05 | 0.05 | 0.05 | 0.05 | 0.05 |
| FB2       | 0.05 | 0.05 | 0.05 | 0.05 | 0.05 | 0.05 | 0.05 | 0.05 | 0.05 | 0.05 | 0.05 |
| FB4       | 0.05 | 0.05 | 0.05 | 0.05 | 0.05 | 0.05 | 0.05 | 0.05 | 0.05 | 0.05 | 0.05 |
| FB6       | 0.05 | 0.05 | 0.05 | 0.05 | 0.05 | 0.05 | 0.05 | 0.05 | 0.05 | 0.05 | 0.05 |
| FB8       | 0.05 | 0.05 | 0.05 | 0.05 | 0.05 | 0.05 | 0.05 | 0.05 | 0.05 | 0.05 | 0.05 |
| FB2 adj   | 0.05 | 0.05 | 0.05 | 0.05 | 0.05 | 0.05 | 0.05 | 0.05 | 0.05 | 0.05 | 0.05 |
| FB4 adj   | 0.05 | 0.05 | 0.05 | 0.05 | 0.05 | 0.05 | 0.05 | 0.05 | 0.05 | 0.05 | 0.05 |
| FB6 adj   | 0.05 | 0.05 | 0.05 | 0.05 | 0.05 | 0.05 | 0.05 | 0.05 | 0.05 | 0.05 | 0.05 |
| FB8 adj   | 0.05 | 0.05 | 0.05 | 0.05 | 0.05 | 0.05 | 0.05 | 0.05 | 0.05 | 0.05 | 0.05 |

Table S1: Sharp Null: For each randomization scheme (rows) and finite population size (columns), the average Type I error across populations of the same size and outcome distribution.

|            | 16   | 24   | 32   | 46   | 64   | 92   | 128  | 182  | 256  | 512  | 1024 |
|-----------:|------|------|------|------|------|------|------|------|------|------|------|
| SR         | 0.85 | 0.92 | 0.94 | 0.95 | 0.95 | 0.95 | 0.95 | 0.95 | 0.95 | 0.95 | 0.95 |
| CR         | 0.84 | 0.92 | 0.94 | 0.95 | 0.95 | 0.95 | 0.95 | 0.95 | 0.95 | 0.95 | 0.95 |
| BS2        | 0.14 | 0.18 | 0.21 | 0.24 | 0.28 | 0.33 | 0.38 | 0.45 | 0.51 | 0.64 | 0.76 |
| BS3        | 0.23 | 0.27 | 0.30 | 0.34 | 0.40 | 0.46 | 0.51 | 0.57 | 0.63 | 0.75 | 0.84 |
| BS4        | 0.36 | 0.39 | 0.41 | 0.46 | 0.50 | 0.56 | 0.61 | 0.66 | 0.73 | 0.82 | 0.88 |
| BS2 adj 1  | 0.24 | 0.31 | 0.35 | 0.41 | 0.47 | 0.53 | 0.59 | 0.65 | 0.71 | 0.82 | 0.88 |
| BS3 adj 1  | 0.35 | 0.45 | 0.52 | 0.60 | 0.67 | 0.73 | 0.77 | 0.82 | 0.85 | 0.90 | 0.93 |
| BS4 adj 1  | 0.51 | 0.60 | 0.67 | 0.73 | 0.78 | 0.83 | 0.86 | 0.88 | 0.90 | 0.93 | 0.94 |
| BS2 adj 2  | 0.24 | 0.26 | 0.27 | 0.29 | 0.32 | 0.35 | 0.41 | 0.46 | 0.52 | 0.64 | 0.76 |
| BS3 adj 2  | 0.38 | 0.50 | 0.54 | 0.54 | 0.54 | 0.56 | 0.58 | 0.62 | 0.66 | 0.76 | 0.84 |
| BS4 adj 2  | 0.51 | 0.63 | 0.72 | 0.76 | 0.75 | 0.75 | 0.75 | 0.75 | 0.78 | 0.84 | 0.89 |
| BS2 adj 3  | 0.18 | 0.22 | 0.25 | 0.30 | 0.34 | 0.41 | 0.46 | 0.53 | 0.60 | 0.72 | 0.82 |
| BS3 adj 3  | 0.25 | 0.31 | 0.36 | 0.41 | 0.47 | 0.53 | 0.59 | 0.65 | 0.71 | 0.81 | 0.88 |
| BS4 adj 3  | 0.39 | 0.44 | 0.48 | 0.53 | 0.58 | 0.64 | 0.69 | 0.74 | 0.80 | 0.87 | 0.91 |
| BS2 adj 4  | 0.21 | 0.21 | 0.23 | 0.26 | 0.30 | 0.34 | 0.39 | 0.44 | 0.52 | 0.64 | 0.76 |
| BS3 adj 4  | 0.41 | 0.52 | 0.58 | 0.65 | 0.70 | 0.75 | 0.79 | 0.83 | 0.86 | 0.90 | 0.93 |
| BS4 adj 4  | 0.46 | 0.48 | 0.49 | 0.52 | 0.55 | 0.59 | 0.63 | 0.69 | 0.74 | 0.82 | 0.88 |
| FB2        | 0.06 | 0.09 | 0.10 | 0.13 | 0.16 | 0.19 | 0.22 | 0.27 | 0.31 | 0.43 | 0.56 |
| FB4        | 0.13 | 0.16 | 0.20 | 0.24 | 0.28 | 0.33 | 0.38 | 0.44 | 0.51 | 0.64 | 0.75 |
| FB6        | 0.19 | 0.23 | 0.27 | 0.31 | 0.36 | 0.42 | 0.48 | 0.54 | 0.61 | 0.73 | 0.83 |
| FB8        | 0.25 | 0.29 | 0.32 | 0.37 | 0.42 | 0.49 | 0.54 | 0.61 | 0.66 | 0.78 | 0.86 |
| FB2 adj    | 0.25 | 0.66 | 0.84 | 0.92 | 0.94 | 0.95 | 0.95 | 0.95 | 0.95 | 0.95 | 0.95 |
| FB4 adj    | 0.49 | 0.82 | 0.90 | 0.94 | 0.95 | 0.95 | 0.95 | 0.95 | 0.95 | 0.95 | 0.95 |
| FB6 adj    | 0.64 | 0.87 | 0.92 | 0.94 | 0.94 | 0.95 | 0.95 | 0.95 | 0.95 | 0.95 | 0.95 |
| FB8 adj    | 0.71 | 0.89 | 0.93 | 0.94 | 0.95 | 0.94 | 0.95 | 0.95 | 0.95 | 0.95 | 0.95 |

Table S2: Sharp Null: For each randomization scheme (rows) and finite population size (columns), the proportion of finite populations (out of 30K) with a Type I error within 0.05 +/- 1.96 * sqrt(0.05 * 0.95 / 10,000).

## S3.2 Normal, 0 Null: Average and convergence ($n = N$)

|           | 16   | 24   | 32   | 46   | 64   | 92   | 128  | 182  | 256  | 512  | 1024 |
|-----------|------|------|------|------|------|------|------|------|------|------|------|
| SR        | 0.09 | 0.08 | 0.07 | 0.06 | 0.06 | 0.06 | 0.06 | 0.05 | 0.05 | 0.05 | 0.05 |
| CR        | 0.09 | 0.08 | 0.07 | 0.06 | 0.06 | 0.06 | 0.05 | 0.05 | 0.05 | 0.05 | 0.05 |
| BS2       | 0.07 | 0.06 | 0.06 | 0.06 | 0.05 | 0.05 | 0.05 | 0.05 | 0.05 | 0.05 | 0.05 |
| BS3       | 0.07 | 0.06 | 0.06 | 0.06 | 0.05 | 0.05 | 0.05 | 0.05 | 0.05 | 0.05 | 0.05 |
| BS4       | 0.08 | 0.07 | 0.06 | 0.06 | 0.06 | 0.05 | 0.05 | 0.05 | 0.05 | 0.05 | 0.05 |
| BS2 adj 1 | 0.08 | 0.07 | 0.06 | 0.06 | 0.06 | 0.05 | 0.05 | 0.05 | 0.05 | 0.05 | 0.05 |
| BS3 adj 1 | 0.08 | 0.07 | 0.06 | 0.06 | 0.06 | 0.05 | 0.05 | 0.05 | 0.05 | 0.05 | 0.05 |
| BS4 adj 1 | 0.09 | 0.07 | 0.06 | 0.06 | 0.06 | 0.05 | 0.05 | 0.05 | 0.05 | 0.05 | 0.05 |
| BS2 adj 2 | 0.08 | 0.07 | 0.06 | 0.06 | 0.05 | 0.05 | 0.05 | 0.05 | 0.05 | 0.05 | 0.05 |
| BS3 adj 2 | 0.08 | 0.07 | 0.06 | 0.06 | 0.06 | 0.05 | 0.05 | 0.05 | 0.05 | 0.05 | 0.05 |
| BS4 adj 2 | 0.09 | 0.07 | 0.06 | 0.06 | 0.06 | 0.05 | 0.05 | 0.05 | 0.05 | 0.05 | 0.05 |
| BS2 adj 3 | 0.07 | 0.06 | 0.06 | 0.06 | 0.05 | 0.05 | 0.05 | 0.05 | 0.05 | 0.05 | 0.05 |
| BS3 adj 3 | 0.07 | 0.06 | 0.06 | 0.06 | 0.05 | 0.05 | 0.05 | 0.05 | 0.05 | 0.05 | 0.05 |
| BS4 adj 3 | 0.08 | 0.06 | 0.06 | 0.06 | 0.05 | 0.05 | 0.05 | 0.05 | 0.05 | 0.05 | 0.05 |
| BS2 adj 4 | 0.07 | 0.06 | 0.06 | 0.06 | 0.05 | 0.05 | 0.05 | 0.05 | 0.05 | 0.05 | 0.05 |
| BS3 adj 4 | 0.06 | 0.06 | 0.06 | 0.05 | 0.05 | 0.05 | 0.05 | 0.05 | 0.05 | 0.05 | 0.05 |
| BS4 adj 4 | 0.07 | 0.06 | 0.06 | 0.06 | 0.05 | 0.05 | 0.05 | 0.05 | 0.05 | 0.05 | 0.05 |
| FB2       | 0.07 | 0.06 | 0.06 | 0.06 | 0.05 | 0.05 | 0.05 | 0.05 | 0.05 | 0.05 | 0.05 |
| FB4       | 0.07 | 0.06 | 0.06 | 0.06 | 0.05 | 0.05 | 0.05 | 0.05 | 0.05 | 0.05 | 0.05 |
| FB6       | 0.07 | 0.06 | 0.06 | 0.06 | 0.05 | 0.05 | 0.05 | 0.05 | 0.05 | 0.05 | 0.05 |
| FB8       | 0.07 | 0.06 | 0.06 | 0.06 | 0.05 | 0.05 | 0.05 | 0.05 | 0.05 | 0.05 | 0.05 |
| FB2 adj   | 0.05 | 0.05 | 0.05 | 0.05 | 0.05 | 0.05 | 0.05 | 0.05 | 0.05 | 0.05 | 0.05 |
| FB4 adj   | 0.07 | 0.06 | 0.06 | 0.06 | 0.05 | 0.05 | 0.05 | 0.05 | 0.05 | 0.05 | 0.05 |
| FB6 adj   | 0.07 | 0.06 | 0.06 | 0.06 | 0.05 | 0.05 | 0.05 | 0.05 | 0.05 | 0.05 | 0.05 |
| FB8 adj   | 0.07 | 0.06 | 0.06 | 0.06 | 0.05 | 0.05 | 0.05 | 0.05 | 0.05 | 0.05 | 0.05 |

Table S3: Normal, 0 Null: For each randomization scheme (rows) and finite population size (columns), the average Type I error across populations of the same size and outcome distribution.

|          | 16   | 24   | 32   | 46   | 64   | 92   | 128  | 182  | 256  | 512  | 1024 |
|---------:|------|------|------|------|------|------|------|------|------|------|------|
| SR       | 0.22 | 0.09 | 0.07 | 0.06 | 0.05 | 0.04 | 0.04 | 0.04 | 0.04 | 0.04 | 0.04 |
| CR       | 0.21 | 0.09 | 0.07 | 0.05 | 0.05 | 0.04 | 0.04 | 0.04 | 0.04 | 0.03 | 0.04 |
| BS2      | 0.06 | 0.05 | 0.05 | 0.04 | 0.04 | 0.04 | 0.04 | 0.04 | 0.04 | 0.04 | 0.04 |
| BS3      | 0.06 | 0.05 | 0.04 | 0.04 | 0.04 | 0.04 | 0.04 | 0.04 | 0.04 | 0.04 | 0.04 |
| BS4      | 0.08 | 0.06 | 0.05 | 0.04 | 0.04 | 0.04 | 0.04 | 0.04 | 0.03 | 0.04 | 0.04 |
| BS2 adj 1 | 0.11 | 0.09 | 0.07 | 0.06 | 0.05 | 0.05 | 0.05 | 0.05 | 0.05 | 0.05 | 0.04 |
| BS3 adj 1 | 0.12 | 0.09 | 0.07 | 0.06 | 0.05 | 0.05 | 0.05 | 0.05 | 0.05 | 0.04 | 0.04 |
| BS4 adj 1 | 0.15 | 0.08 | 0.07 | 0.05 | 0.05 | 0.05 | 0.04 | 0.04 | 0.04 | 0.04 | 0.04 |
| BS2 adj 2 | 0.11 | 0.07 | 0.06 | 0.05 | 0.04 | 0.04 | 0.04 | 0.04 | 0.04 | 0.04 | 0.04 |
| BS3 adj 2 | 0.13 | 0.08 | 0.06 | 0.05 | 0.05 | 0.04 | 0.04 | 0.04 | 0.04 | 0.04 | 0.04 |
| BS4 adj 2 | 0.15 | 0.08 | 0.07 | 0.05 | 0.05 | 0.04 | 0.04 | 0.04 | 0.04 | 0.04 | 0.04 |
| BS2 adj 3 | 0.07 | 0.06 | 0.05 | 0.05 | 0.05 | 0.05 | 0.05 | 0.05 | 0.05 | 0.04 | 0.04 |
| BS3 adj 3 | 0.07 | 0.05 | 0.05 | 0.05 | 0.04 | 0.04 | 0.04 | 0.04 | 0.04 | 0.04 | 0.04 |
| BS4 adj 3 | 0.08 | 0.06 | 0.05 | 0.05 | 0.05 | 0.04 | 0.04 | 0.04 | 0.04 | 0.04 | 0.04 |
| BS2 adj 4 | 0.13 | 0.15 | 0.14 | 0.11 | 0.10 | 0.09 | 0.09 | 0.08 | 0.08 | 0.08 | 0.07 |
| BS3 adj 4 | 0.13 | 0.09 | 0.08 | 0.07 | 0.07 | 0.06 | 0.06 | 0.06 | 0.06 | 0.06 | 0.06 |
| BS4 adj 4 | 0.15 | 0.16 | 0.14 | 0.11 | 0.10 | 0.09 | 0.08 | 0.08 | 0.08 | 0.08 | 0.07 |
| FB2      | 0.05 | 0.05 | 0.05 | 0.04 | 0.04 | 0.04 | 0.04 | 0.04 | 0.04 | 0.03 | 0.04 |
| FB4      | 0.05 | 0.05 | 0.04 | 0.04 | 0.04 | 0.04 | 0.04 | 0.04 | 0.04 | 0.03 | 0.04 |
| FB6      | 0.06 | 0.05 | 0.05 | 0.04 | 0.04 | 0.04 | 0.04 | 0.04 | 0.04 | 0.03 | 0.04 |
| FB8      | 0.05 | 0.05 | 0.05 | 0.04 | 0.04 | 0.04 | 0.04 | 0.04 | 0.03 | 0.04 | 0.04 |
| FB2 adj  | 0.04 | 0.04 | 0.04 | 0.04 | 0.04 | 0.04 | 0.04 | 0.04 | 0.03 | 0.03 | 0.04 |
| FB4 adj  | 0.05 | 0.05 | 0.04 | 0.04 | 0.04 | 0.04 | 0.04 | 0.04 | 0.04 | 0.03 | 0.04 |
| FB6 adj  | 0.06 | 0.05 | 0.05 | 0.04 | 0.04 | 0.04 | 0.04 | 0.04 | 0.04 | 0.03 | 0.04 |
| FB8 adj  | 0.05 | 0.05 | 0.04 | 0.04 | 0.04 | 0.04 | 0.04 | 0.04 | 0.04 | 0.03 | 0.04 |

Table S4: Normal, 0 Null: For each randomization scheme (rows) and finite population size (columns), the proportion of finite populations (out of 30K) with a Type I error within 0.05 +/- 1.96 * sqrt(0.05 * 0.95 / 10,000).

## S3.3 Normal, Normal Null Type I error: Average and convergence ($n = N$)

|         | 16   | 24   | 32   | 46   | 64   | 92   | 128  | 182  | 256  | 512  | 1024 |
|---------|------|------|------|------|------|------|------|------|------|------|------|
| SR      | 0.05 | 0.05 | 0.05 | 0.05 | 0.05 | 0.05 | 0.05 | 0.05 | 0.05 | 0.05 | 0.05 |
| CR      | 0.05 | 0.05 | 0.05 | 0.05 | 0.05 | 0.05 | 0.05 | 0.05 | 0.05 | 0.05 | 0.05 |
| BS2     | 0.05 | 0.05 | 0.05 | 0.05 | 0.05 | 0.05 | 0.05 | 0.05 | 0.05 | 0.05 | 0.05 |
| BS3     | 0.05 | 0.05 | 0.05 | 0.05 | 0.05 | 0.05 | 0.05 | 0.05 | 0.05 | 0.05 | 0.05 |
| BS4     | 0.05 | 0.05 | 0.05 | 0.05 | 0.05 | 0.05 | 0.05 | 0.05 | 0.05 | 0.05 | 0.05 |
| BS2 adj 1 | 0.05 | 0.05 | 0.05 | 0.05 | 0.05 | 0.05 | 0.05 | 0.05 | 0.05 | 0.05 | 0.05 |
| BS3 adj 1 | 0.05 | 0.05 | 0.05 | 0.05 | 0.05 | 0.05 | 0.05 | 0.05 | 0.05 | 0.05 | 0.05 |
| BS4 adj 1 | 0.05 | 0.05 | 0.05 | 0.05 | 0.05 | 0.05 | 0.05 | 0.05 | 0.05 | 0.05 | 0.05 |
| BS2 adj 2 | 0.05 | 0.05 | 0.05 | 0.05 | 0.05 | 0.05 | 0.05 | 0.05 | 0.05 | 0.05 | 0.05 |
| BS3 adj 2 | 0.05 | 0.05 | 0.05 | 0.05 | 0.05 | 0.05 | 0.05 | 0.05 | 0.05 | 0.05 | 0.05 |
| BS4 adj 2 | 0.05 | 0.05 | 0.05 | 0.05 | 0.05 | 0.05 | 0.05 | 0.05 | 0.05 | 0.05 | 0.05 |
| BS2 adj 3 | 0.05 | 0.05 | 0.05 | 0.05 | 0.05 | 0.05 | 0.05 | 0.05 | 0.05 | 0.05 | 0.05 |
| BS3 adj 3 | 0.05 | 0.05 | 0.05 | 0.05 | 0.05 | 0.05 | 0.05 | 0.05 | 0.05 | 0.05 | 0.05 |
| BS4 adj 3 | 0.05 | 0.05 | 0.05 | 0.05 | 0.05 | 0.05 | 0.05 | 0.05 | 0.05 | 0.05 | 0.05 |
| BS2 adj 4 | 0.05 | 0.05 | 0.05 | 0.05 | 0.05 | 0.05 | 0.05 | 0.05 | 0.05 | 0.05 | 0.05 |
| BS3 adj 4 | 0.05 | 0.05 | 0.05 | 0.05 | 0.05 | 0.05 | 0.05 | 0.05 | 0.05 | 0.05 | 0.05 |
| BS4 adj 4 | 0.05 | 0.05 | 0.05 | 0.05 | 0.05 | 0.05 | 0.05 | 0.05 | 0.05 | 0.05 | 0.05 |
| FB2     | 0.05 | 0.05 | 0.05 | 0.05 | 0.05 | 0.05 | 0.05 | 0.05 | 0.05 | 0.05 | 0.05 |
| FB4     | 0.05 | 0.05 | 0.05 | 0.05 | 0.05 | 0.05 | 0.05 | 0.05 | 0.05 | 0.05 | 0.05 |
| FB6     | 0.05 | 0.05 | 0.05 | 0.05 | 0.05 | 0.05 | 0.05 | 0.05 | 0.05 | 0.05 | 0.05 |
| FB8     | 0.05 | 0.05 | 0.05 | 0.05 | 0.05 | 0.05 | 0.05 | 0.05 | 0.05 | 0.05 | 0.05 |
| FB2 adj | 0.05 | 0.05 | 0.05 | 0.05 | 0.05 | 0.05 | 0.05 | 0.05 | 0.05 | 0.05 | 0.05 |
| FB4 adj | 0.05 | 0.05 | 0.05 | 0.05 | 0.05 | 0.05 | 0.05 | 0.05 | 0.05 | 0.05 | 0.05 |
| FB6 adj | 0.05 | 0.05 | 0.05 | 0.05 | 0.05 | 0.05 | 0.05 | 0.05 | 0.05 | 0.05 | 0.05 |
| FB8 adj | 0.05 | 0.05 | 0.05 | 0.05 | 0.05 | 0.05 | 0.05 | 0.05 | 0.05 | 0.05 | 0.05 |

Table S5: Normal, Normal Null: For each randomization scheme (rows) and finite population size (columns), the average Type I error across populations of the same size and outcome distribution.

|            | 16   | 24   | 32   | 46   | 64   | 92   | 128  | 182  | 256  | 512  | 1024 |
|------------|------|------|------|------|------|------|------|------|------|------|------|
| SR         | 0.04 | 0.04 | 0.04 | 0.04 | 0.04 | 0.04 | 0.04 | 0.04 | 0.04 | 0.04 | 0.03 |
| CR         | 0.04 | 0.04 | 0.04 | 0.04 | 0.04 | 0.04 | 0.04 | 0.04 | 0.04 | 0.04 | 0.03 |
| BS2        | 0.04 | 0.04 | 0.04 | 0.04 | 0.04 | 0.04 | 0.04 | 0.03 | 0.04 | 0.04 | 0.04 |
| BS3        | 0.04 | 0.04 | 0.04 | 0.03 | 0.04 | 0.04 | 0.04 | 0.04 | 0.04 | 0.04 | 0.04 |
| BS4        | 0.04 | 0.04 | 0.04 | 0.04 | 0.04 | 0.04 | 0.04 | 0.04 | 0.04 | 0.04 | 0.03 |
| BS2 adj 1  | 0.05 | 0.05 | 0.05 | 0.04 | 0.05 | 0.04 | 0.04 | 0.04 | 0.04 | 0.04 | 0.04 |
| BS3 adj 1  | 0.05 | 0.05 | 0.05 | 0.04 | 0.04 | 0.04 | 0.04 | 0.04 | 0.04 | 0.04 | 0.04 |
| BS4 adj 1  | 0.05 | 0.05 | 0.04 | 0.04 | 0.04 | 0.04 | 0.04 | 0.04 | 0.04 | 0.04 | 0.04 |
| BS2 adj 2  | 0.05 | 0.05 | 0.04 | 0.04 | 0.04 | 0.04 | 0.04 | 0.04 | 0.04 | 0.04 | 0.03 |
| BS3 adj 2  | 0.05 | 0.05 | 0.04 | 0.04 | 0.04 | 0.04 | 0.04 | 0.04 | 0.04 | 0.04 | 0.03 |
| BS4 adj 2  | 0.05 | 0.05 | 0.04 | 0.04 | 0.04 | 0.04 | 0.04 | 0.04 | 0.04 | 0.04 | 0.03 |
| BS2 adj 3  | 0.05 | 0.05 | 0.05 | 0.05 | 0.05 | 0.05 | 0.05 | 0.05 | 0.05 | 0.05 | 0.04 |
| BS3 adj 3  | 0.05 | 0.04 | 0.04 | 0.04 | 0.04 | 0.04 | 0.04 | 0.04 | 0.04 | 0.04 | 0.04 |
| BS4 adj 3  | 0.05 | 0.04 | 0.04 | 0.04 | 0.04 | 0.04 | 0.04 | 0.04 | 0.04 | 0.04 | 0.04 |
| BS2 adj 4  | 0.10 | 0.10 | 0.10 | 0.10 | 0.09 | 0.08 | 0.08 | 0.08 | 0.08 | 0.08 | 0.08 |
| BS3 adj 4  | 0.09 | 0.07 | 0.07 | 0.06 | 0.06 | 0.06 | 0.06 | 0.06 | 0.06 | 0.06 | 0.06 |
| BS4 adj 4  | 0.12 | 0.10 | 0.10 | 0.09 | 0.08 | 0.08 | 0.08 | 0.08 | 0.08 | 0.08 | 0.08 |
| FB2        | 0.04 | 0.04 | 0.04 | 0.04 | 0.03 | 0.03 | 0.04 | 0.04 | 0.04 | 0.03 | 0.04 |
| FB4        | 0.04 | 0.04 | 0.04 | 0.03 | 0.04 | 0.04 | 0.04 | 0.04 | 0.04 | 0.04 | 0.03 |
| FB6        | 0.04 | 0.04 | 0.04 | 0.04 | 0.04 | 0.03 | 0.04 | 0.04 | 0.04 | 0.04 | 0.04 |
| FB8        | 0.04 | 0.04 | 0.04 | 0.04 | 0.04 | 0.04 | 0.04 | 0.04 | 0.04 | 0.04 | 0.03 |
| FB2 adj    | 0.04 | 0.04 | 0.04 | 0.04 | 0.04 | 0.04 | 0.04 | 0.04 | 0.04 | 0.04 | 0.04 |
| FB4 adj    | 0.04 | 0.04 | 0.04 | 0.04 | 0.04 | 0.04 | 0.04 | 0.04 | 0.04 | 0.04 | 0.03 |
| FB6 adj    | 0.04 | 0.04 | 0.04 | 0.04 | 0.04 | 0.04 | 0.04 | 0.04 | 0.04 | 0.04 | 0.04 |
| FB8 adj    | 0.04 | 0.04 | 0.04 | 0.04 | 0.04 | 0.04 | 0.04 | 0.04 | 0.04 | 0.04 | 0.03 |

Table S6: Normal, Normal Null: For each randomization scheme (rows) and finite population size (columns), the proportion of finite populations (out of 30K) with a Type I error within 0.05 +/- 1.96 * sqrt(0.05 * 0.95 / 10,000).